\documentclass[11pt,a4paper]{article}

\usepackage[utf8]{inputenc}
\usepackage[T1]{fontenc}
\usepackage{lmodern}

\usepackage{amsmath,amssymb,amsthm,mathtools,physics}
\usepackage{bm,enumerate,mathrsfs,tensor,upgreek,xcolor,enumitem}
\usepackage{booktabs}
\usepackage[a4paper,margin=1in]{geometry}
\usepackage{authblk}

\newcommand{\CV}{\mathcal{V}}

\title{Turbulent Pair Dispersion with Stochastic Generative Diffusion Models}

\author[2]{Andrei Pantea}
\author[1]{Luca Biferale}
\author[1]{Michele Buzzicotti}
\author[2]{Guillaume Charpiat}
\author[2]{Sergio Chibbaro}
\author[1]{Tianyi Li}

\affil[1]{Department of Physics and INFN, University of Rome ``Tor Vergata'', Via della Ricerca Scientifica 1, 00133 Rome, Italy}
\affil[2]{Universit\'e Paris-Saclay, CNRS, INRIA, LISN, UMR 9015, F-91405 Orsay Cedex, France}

\date{} 

\begin{document}

\maketitle

\abstract{Recent advances in data-driven modeling have shown that diffusion models can successfully generate synthetic Lagrangian trajectories in turbulent flows. Building on this progress, we extend the method to the joint generation of pairs of Lagrangian velocity trajectories, enabling a fully data-driven representation of turbulent pair dispersion, a long-standing fundamental problem with broad relevance in fluid dynamics.
We demonstrate that diffusion models accurately reproduce the evolution of particle-pair separation, including deviations from Richardson’s classical scaling law, while simultaneously preserving all key single-particle statistical properties reported in previous studies. These findings underscore the potential of diffusion-based generative models to emulate high-dimensional, multi-scale turbulent dynamics, further establishing them as a powerful tool for scientific modeling and for future geophysical and astrophysical applications.}

\section{Introduction}
\label{sect:intro}
At high Reynolds numbers, molecular diffusion makes a negligible contribution to spatial transport~\cite{pope2001turbulent}. Modeling how turbulence redistributes particles is thus of fundamental importance for understanding how turbulent motions distribute scalar fields such as temperature, humidity, salinity, and any passive chemical species or concentration, with crucial importance across fluid dynamics, geophysics, and environmental sciences~\cite{sawford2001turbulent,falkovich2001particles,salazar2009two,bourgoin2006role,toschi2009lagrangian,mathai2018dispersion,sreenivasan2019turbulent,shnapp2023universal}.

\begin{figure*}[t]
\includegraphics[width=1.\textwidth]{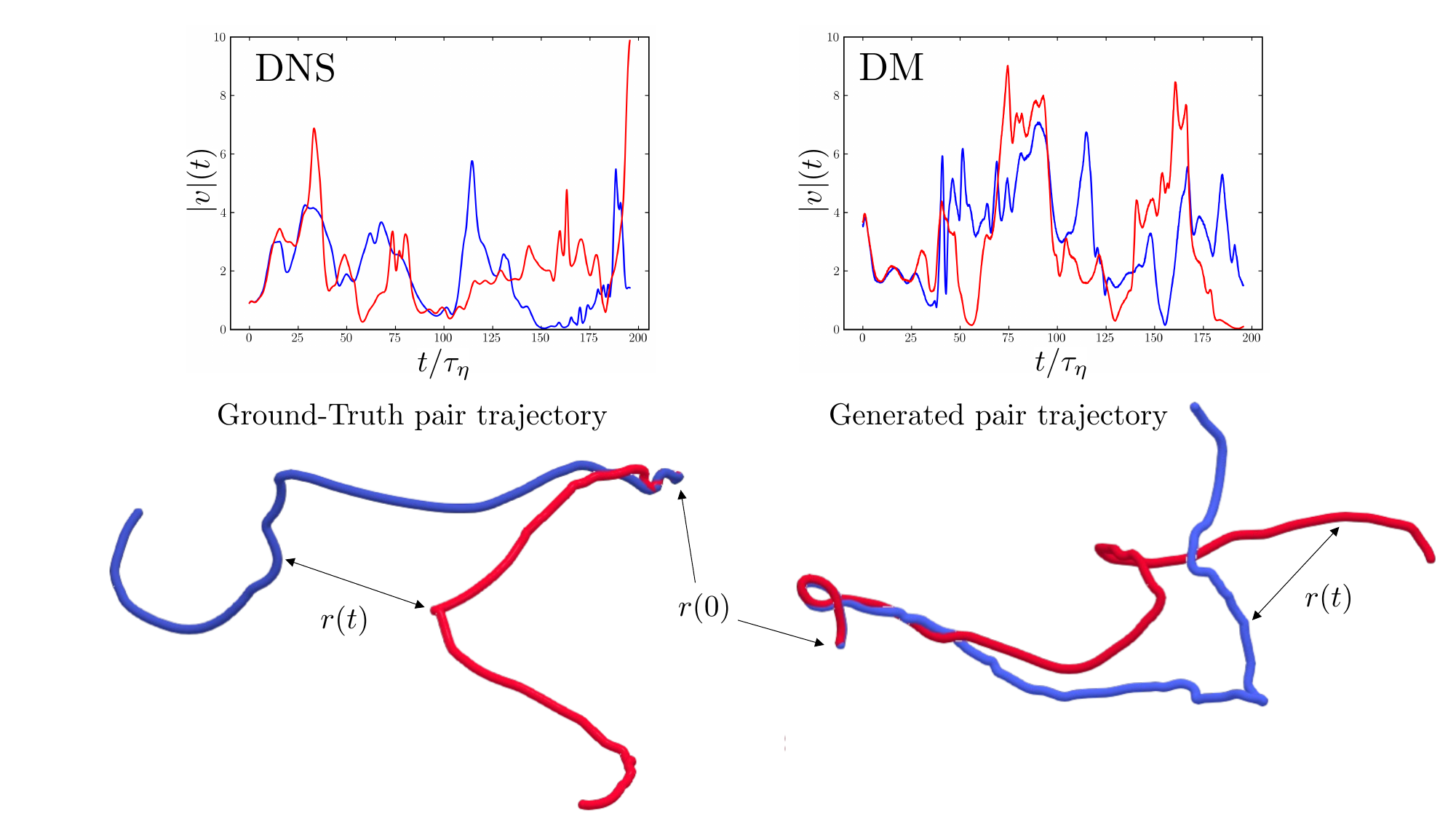}
\caption{Comparison between direct numerical simulation (DNS) and diffusion model (DM) generated particle pairs. Top: time evolution of the velocity magnitude $|\mathbf{V}(t)|$ for the two particles forming a pair, shown for DNS (left) and DM (right). 
Bottom: corresponding three-dimensional trajectories of the two particles (red and blue curves), initialized at separation $r(0)$. The instantaneous inter-particle distance at a fixed time $t$ during the evolution is denoted as $r(t)$.}
\label{fig:fig_1}
\end{figure*}
\noindent
The paradigmatic problem of pairs separation of particles advected by fully developed turbulent flows was first described by Richardson~\cite{richardson1926atmospheric,davidson2011voyage} one hundred years ago, 1926.
In his framework, 
considering the diffusive effect exerted by the turbulent motion, Richardson argued that the time evolution of the distance neighbor function could be described by a  diffusion equation for the probability density function of distances between couples of particles at a given time, 
\begin{equation}
    \partial_t P(r,t) = \frac{1}{r^{2}} \partial_r \left[ r^{2} D(r) \, \partial_r P(r,t) \right].
\end{equation}
From experimental data, Richardson proposed his celebrated “4/3” law $    D(r) \sim  r^{4/3}$.
The same scaling can be derived by simple dimensional argument as suggested by Oboukhov~\cite{monin2013statistical} considering that $\langle r^{2}(t) \rangle$ evolves according to a scale-dependent diffusivity, $D(r)$,
\begin{equation}
    \frac{d\langle r^{2}(t)\rangle}{dt} = D(r),
\end{equation}
where $r(t)$ is the instantaneous inter-particle distance.  Invoking Kolmogorov’s 1941 (K41) scaling for inertial-range velocity increments~\cite{frisch1996turbulence}, it is obtained~\cite{monin2013statistical}
\begin{equation}
    D(r) \sim \varepsilon^{1/3} r^{4/3},
\end{equation}
with $\varepsilon$ the mean kinetic energy dissipation rate.  
Substituting the Richardson-Kolmogorov scaling  $D(r) \sim \varepsilon^{1/3} r^{4/3}$ leads to the well-known Richardson PDF,
\begin{equation}
    P(r,t) \sim \frac{r^2}{(\varepsilon^{1/3} t)^{9/2}} 
    \exp\!\left[-\,C \frac{r^{2/3}}{\varepsilon^{1/3} t} \right],
    \label{eq:pdf_rich}
\end{equation}
which predicts a stretched-exponential distribution of separations in the inertial range.
This relation yields the celebrated Richardson $t^{3}$ law for pair dispersion~\cite{monin2013statistical,majda1999simplified,zouari1994derivation,boffetta1999pair,boffetta2002relative,boffetta2002statistics,espanol2025effect},
\begin{equation}
    \langle r^{2}(t)\rangle \sim \varepsilon\, t^{3},
    \label{eq:5}
\end{equation}
a super-diffusive prediction reflecting the rapid acceleration of particle separation as they encounter progressively larger turbulent eddies~\cite{bourgoin2015turbulent,falkovich2001particles}.
Despite its foundational role, it is now well established by a number of experimental and numerical studies, that Richardson’s predictions are not strictly valid in real turbulent flows~\cite{novikov1989two,grossmann1984unified,crisanti1987multifractal}.  
While the Richardson law Eq. (\ref{eq:5}) is not affected by intermittency corrections~\cite{boffetta1999pair,boffetta2002relative,boffetta2002statistics}, higher moments in general are ~\cite{scatamacchia2012extreme}.
Deviations arise from several physical mechanisms, including:  
(i) the non-Markovian nature of temporal correlations in the flow, which undermine the assumption of scale-local and memory-less, time independent, diffusion;  
(ii) strong non-Gaussian fluctuations of turbulent velocity increments, especially at small and intermediate/inertial scales which violates the K41 scaling; and  
(iii) finite-Reynolds-number effects, which limit the extent of the inertial range and modify the scaling behavior of separation statistics.  
These factors collectively lead to measurable departures from the idealized Richardson picture in both experiments and numerical simulations~\cite{li2024relative,scatamacchia2012extreme,bec2010turbulent}. The chaotic, multiscale nature of turbulence challenges both dispersion modeling and tracking, as even small separations quickly diverge~\cite{calascibetta2023optimal,calascibetta2023taming}.

Notwithstanding the extensive literature on turbulent dispersion, a stochastic model capable of modeling realistic particle-pair trajectories across all relevant turbulent scales is still lacking.  
Existing approaches typically capture only limited regimes or rely on simplified assumptions that prevent them from reproducing the full multiscale, intermittent nature of Lagrangian dynamics in turbulence.

In our recent work~\cite{li2024synthetic}, we addressed this limitation through a data-driven framework based on denoising diffusion probabilistic models (DDPMs)~\cite{sohl2015deep,ho2020denoising}.  
Trained on Lagrangian data from high-resolution direct numerical simulation (DNS) of homogeneous isotropic turbulence (HIT), these models generate Lagrangian velocity trajectories that accurately reproduce high-order statistical properties across the entire range of temporal scales available in the training data.  
They thus offer an efficient alternative to obtaining Lagrangian datasets via DNS or laboratory experiments~\cite{buzzicotti2023data}, substantially reducing the computational and experimental cost of data acquisition.

\begin{figure*}[t]
\includegraphics[width=1.\textwidth]{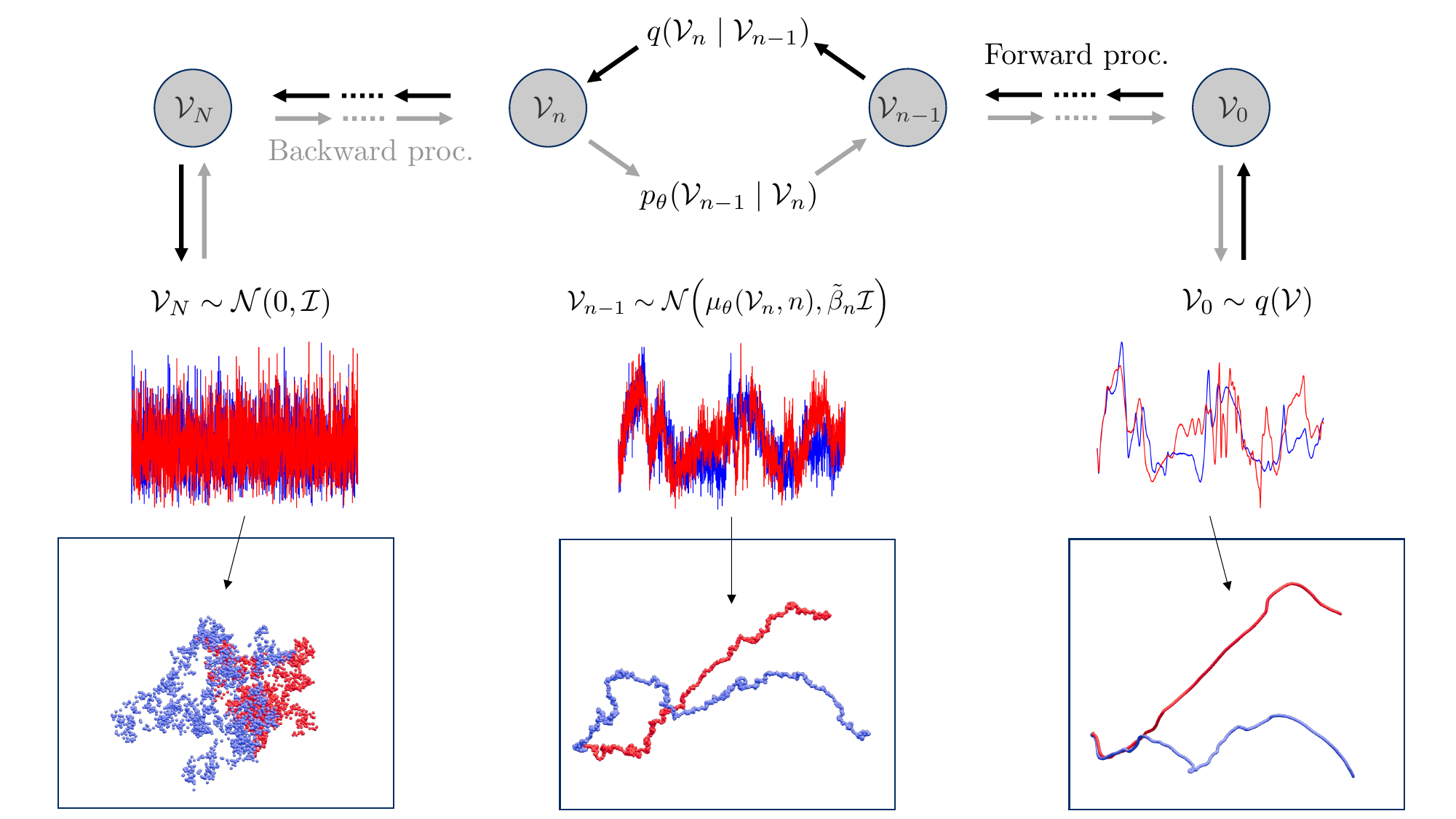}
\caption{Schematic representation of the diffusion model generative framework. 
Top: forward diffusion process $q(\CV_n \mid \CV_{n-1})$, progressively corrupting the clean data sample $\CV_0 \sim q(\CV)$ into Gaussian noise $\CV_N \sim \mathcal{N}(0,\mathcal{I})$ through a sequence of Markovian steps indicated by the black arrows. 
The backward (generative) process $p_\theta(\CV_{n-1} \mid \CV_n)$, inidicated by the gray arrows, starts from pure noise, and inverts the forward transformation by predicting with a trained neural network the mean $\mu_\theta(\CV_n,n)$ of the transition kernel, while the variance $\tilde{\beta}_n I$ is prescribed by the knowledge of the forward process. 
Middle row: illustration of intermediate noisy velocity signals during the forward and backward evolution. 
Bottom row: corresponding reconstruction of particle trajectories, showing the progressive emergence of coherent multiscale structures as the denoising process proceeds from $\CV_N$ to $\CV_0$.}
\label{fig:fig_gen_proc}
\end{figure*}

Building on this foundation, we have shown that the method extends naturally to a variety of physical settings by augmenting the underlying dynamical representation within the training data.  This diffusion-based framework can faithfully generate tracer particles, light and heavy inertial particles, with excellent agreement with reference statistics~\cite{li2024generative}, as well as high-energy charged particles propagating through strong magnetic turbulence~\cite{martin2025generation}.
Recently, we demonstrated that conditioning the generative process to match some observations, allows for the accurate reconstruction of the Lagrangian trajectory, even when only sparse or incomplete observations are available~\cite{li2025stochastic}. This has been demonstrated on DNS and experimental data from oceanic drifters.
This work investigates the possibility of using diffusion models to directly generate particle pairs in turbulence. 
Our goal is to develop a generative process that can reproduce the correct single-particle Lagrangian statistics documented in previous studies and the full dynamical evolution of particle pair separation. 

As illustrated in Fig.~\ref{fig:fig_1}, in this work we aim to model the time evolution of the two-particle velocity signals, from which the corresponding pair trajectories can be reconstructed by temporal integration starting from any initial position. 
In doing so, a successful model must simultaneously capture the intermittent single-particle velocity statistics and the correct multiscale separation dynamics, ensuring consistency between individual Lagrangian behavior and pair dispersion statistics. 
Figure~\ref{fig:fig_1} shows a comparison between a trajectory from the DNS ground-truth data (left) and one generated by the DM model (right).

Before presenting the quantitative comparison of the results, the paper is organized as follows. 
The Methods section describes the training dataset and the DM framework implemented in this work. 
The Results section presents and compares pair and single-particle statistics with those obtained from the DNS training data. 
Finally, the Conclusion summarizes the main findings.

\section{\label{sect:methods} Methods} In this section, we describe the training dataset and detail the diffusion model framework, including network architecture, parameters, and implementation.

\subsection{Training Dataset}
The dataset employed in this study is extracted from the database provided by~\cite{biferale2023turb}, obtained through DNS of homogeneous isotropic turbulence. The simulation tracks a total of $N_p = 327{,}680$ Lagrangian particle trajectories over a time interval $T = 200\,\tau_\eta$, with data stored at temporal increments of $\mathrm{d}t_s = 15\,\mathrm{d}t \approx 0.1\,\tau_\eta$. Consequently, each particle trajectory consists of $K = 2000$ recorded time steps. The flow is simulated in a triply periodic cubic domain of side length $L = 2\pi$, discretized using $1024^3$ collocation points. 
Prior to particle injection, the Eulerian velocity field was evolved until a statistically stationary turbulent regime was achieved. Particles were then introduced at random positions in tuples of four, with initial separations of $dx/2$, where $dx$ denotes the Eulerian resolution. By selecting pairs of trajectories with initially close positions, we obtain the dataset employed for training the generation of pairs of Lagrangian trajectories.
For the purposes of the present analysis, a random subsample of $100{,}000$ trajectory pairs was extracted from the full database.

\subsection{Diffusion Model Framework}
In our notation, each trajectory pair is represented as
$$
    \CV = \{(V_i^1(t_k), V_i^2(t_k)) \mid t_k \in [0,T], \; i=x,y,z\},
$$
where $k = 1,\dots,K$ denotes the discretized sampling times, and $(V^1, V^2)$ corresponds to the two Lagrangian pairs of particles initialized at separation $r(t_0=0) = dx/2$ at initial time. The empirical distribution of the ground-truth trajectories obtained from DNS is denoted by $q(\CV)$.
The DMs framework consists of two processes: the forward and the backward (or reverse) process.\\

\noindent
The \textit{forward} diffusion process consists of $N$ Markovian noising steps, starting from a DNS sample $\CV_0 = \CV$. Each step, $n = 1,\dots,N$, is defined as
\begin{equation}
    q(\CV_n \mid \CV_{n-1}) :\quad 
    \CV_n \sim \mathcal{N}\!\left(\sqrt{1 - \beta_n}\,\CV_{n-1}, \, \beta_n \bm{I}\right),
\end{equation}
meaning that $\CV_n$ is drawn from a Gaussian distribution with mean $\sqrt{1-\beta_n}\CV_{n-1}$ and covariance $\beta_n \bm{I}$. 
The complete forward process can therefore be written as
\begin{equation}
    q(\CV_{1:N} \mid \CV_0)
    \coloneqq \prod_{n=1}^{N} q(\CV_n \mid \CV_{n-1}),
\end{equation}
where $\CV_{1:N}$ denotes the entire sequence of noisy trajectories $\CV_1,\CV_2,\dots,\CV_N$ associated with a single clean input $\CV_0$. 
The schedule $\{\beta_n\}_{n=1}^N$ is predefined, with $N$ sufficiently large so that $\CV_N$ approaches a standard Gaussian distribution, i.e., $\CV_N \sim \mathcal{N}(0,\bm{I})$. 
Noise is injected progressively, first disrupting small-scale correlations and then gradually erasing the large-scale structure of the trajectories.\\
\noindent
Using the cumulative coefficients $\bar{\alpha}_n\coloneqq \prod_{s=1}^{n} \alpha_s$, with $\alpha_n \coloneqq 1 - \beta_n$ the forward process admits a closed-form expression at arbitrary step $n$,
\begin{equation}
\CV_n = \sqrt{\bar{\alpha}_n}\, \CV_0 
+ \sqrt{1 - \bar{\alpha}_n}\, \varepsilon,
\qquad 
\varepsilon \sim \mathcal{N}(0, \bm{I}),
\end{equation}
which allows direct sampling of the noisy trajectory without explicitly iterating over all intermediate diffusion steps.
Following the approach reported in~\cite{li2024synthetic}, we adopt a non-linear variance schedule based on a hyperbolic tangent profile. 
For a total number of diffusion steps $N=400$, the cumulative coefficient $\bar{\alpha}_n$ is prescribed as
\begin{equation}
\bar{\alpha}_n =
\frac{-\tanh\!\left(7n/N - 6\right) + \tanh(1)}
{-\tanh(-6) + \tanh(1)},
\qquad n = 0,\dots,N,
\end{equation}
which ensures $\bar{\alpha}_0 = 1$ and a smooth monotonic decay as $n \to N$, concentrating diffusion steps in the dynamically relevant intermediate regime.\\

\noindent
The \textit{backward} process reverses the forward transformation by learning the conditional transition probabilities $p_\theta(\CV_{n-1}\mid \CV_n)$. 
Starting from Gaussian noise sampled from $p(\CV_N)=\mathcal{N}(\bm{0},\bm{I})$, new trajectory pairs are generated according to
\begin{equation}
    p_\theta(\CV_{0:N})
    = p(\CV_N)\prod_{n=1}^{N} p_\theta(\CV_{n-1}\mid \CV_n).
\end{equation}
In the continuous-diffusion limit, achieved through a suitable choice of the variance schedule and a sufficiently large number of diffusion steps, the backward transition $p_\theta(\CV_{n-1}\mid \CV_n)$ retains a Gaussian form analogous to that of the forward process. 
The neural network is therefore trained to predict the mean $\mu_\theta(\CV_n,n)$, while the variance $\tilde{\beta}_n$ is determined from the forward process~\cite{li2024synthetic}:
\begin{equation}\label{equ:backward}
    p_\theta(\CV_{n-1}\mid \CV_n)
    :\quad \CV_{n-1} \sim 
    \mathcal{N}\!\left(\mu_{\theta}(\CV_n,n),\,
    \tilde{\beta}_n \bm{I} \right).
\end{equation}
Training the neural network consists in minimizing an upper bound on the negative log-likelihood of the data,
\begin{equation}\label{equ:nll}
    \mathbb{E}_{q(\CV_0)}\!\left[-\log\big(p_\theta(\CV_0)\big)\right],
\end{equation}
whose derivation and interpretation are detailed in~\cite{li2024synthetic}. 
Once trained, the backward diffusion progressively reconstructs the structure of the trajectories: large-scale features emerge during the early denoising stages, while small-scale intense fluctuations and smoother regions are recovered in the final steps of the generative process.
The schematics of the forward process and the backward generative mechanisms are illustrated in Fig.~\ref{fig:fig_gen_proc}, highlighting the progressive injection of Gaussian noise during the forward steps, represented by the black arrows starting from a velocity sample, and the subsequent noise reduction performed by the backward generative process, represented by the gray arrows, which starts from pure noise and progressively reconstructs coherent multiscale structures through the learned denoising dynamics.
The UNet architecture implemented in this work is the same as that previously shown to be successful for single-particle generation~\cite{li2024synthetic}. 
The main architectural details and training hyperparameters are summarized in Table~\ref{tab:unet_hyperparameters}.
\begin{table}[h]
\centering
\caption{UNet Hyperparameters.}
\begin{tabular}{lc}
\toprule
\textbf{Hyperparameters} & \textbf{Value} \\
\midrule
Diffusion steps        & 400 \\
Noise schedule         & tanh6-1 \\
Model size             & 417M \\
Channels               & 512 \\
Depth                  & 3 \\
Channels multiple      & 1,1,2,3,4 \\
Heads                  & 4 \\
Attention resolution   & 250,125 \\
Batch size             & 48 \\
Learning rate          & 1e-5 \\
\bottomrule
\end{tabular}
\label{tab:unet_hyperparameters}
\end{table}

\begin{figure}[t]
\includegraphics[width=1.\columnwidth]{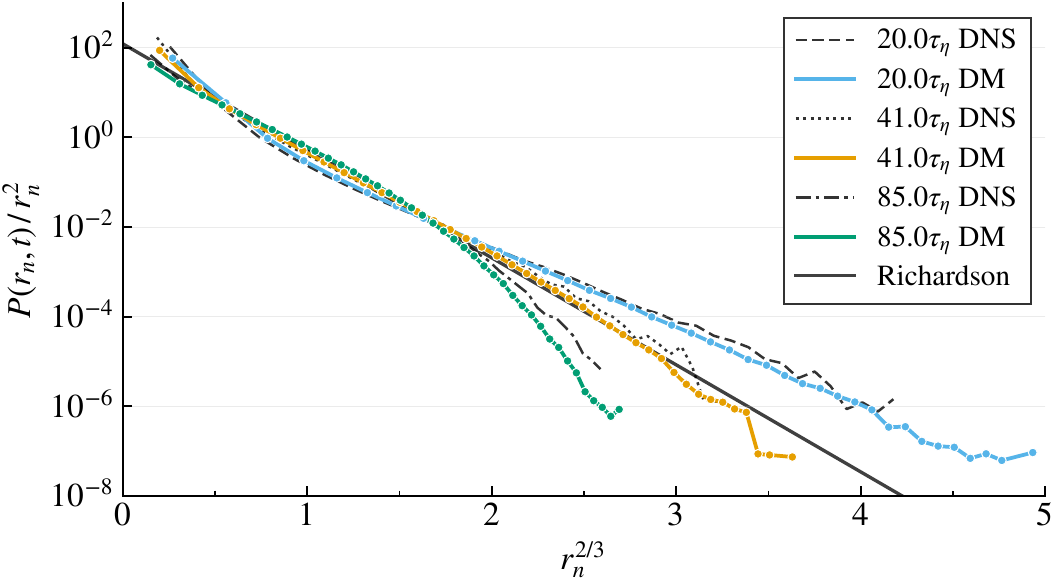}
\caption{Probability density function of pair separation $P(r_n,t)$, compensated by $r_n^2$, plotted as a function of $r_n^{2/3}$ at different times $t$ (in units of $\tau_\eta$). 
Results from DNS (dashed black lines) and the diffusion model (DM) (solid colored lines with symbols) are compared against the Richardson self-similar prediction (solid black line). 
The DM accurately reproduces the full shape of the separation PDF and captures deviations from the ideal Richardson stretched-exponential behavior at finite Reynolds number.}
\label{fig:fig_2}
\end{figure}

\section{Results}
\label{sect:results}

In this section we assess the diffusion model generation over the full dynamical complexity of turbulent pair motion. 
As mentioned above, Figure~\ref{fig:fig_1} shows a representative trajectory pair generated by diffusion model compared to DNS. 
This qualitative visualization highlights the intrinsic difficulty of the generative task: the model must simultaneously learn the six velocity components of the two particles, $(V^1_x(t),V^1_y(t),V^1_z(t))$ and $(V^2_x(t),V^2_y(t),V^2_z(t))$, in such a way that both single-particle statistics and pair-dispersion properties are correctly preserved. 
From the qualitative comparison of the velocity magnitude evolution in the top row of Figure~\ref{fig:fig_1}, we can see that there is an initial transient time, in these examples of the order of $t\sim 20\tau_\eta$, during which the two particle velocities are highly correlated before starting to deviate their evolutions. Both DNS and DM velocities show the coexistence of smooth fluctuations enriched with more intense and rare (intermittent) velocity bursts. 
The resulting particle trajectories are characterized by smooth, almost straight regions and intense vortical-shaped behavior associated with the bursts. Such dynamics are reproduced by the DM without visible artifacts. 
This qualitatively indicates that the denoising generative process consistently reconstructs coherent structures in Lagrangian turbulence.

To quantitatively substantiate these observations, we now present detailed statistical comparisons between DNS and DM data. 
In Fig.~\ref{fig:fig_2}, we analyze pair-dispersion statistics through the PDF, $P(r,t)$, scaled in terms of the variable $r_n(t)=r(t)/(\varepsilon \, t^3)^{1/2}$, at three different times, and compared against the Richardson's prediction~\eqref{eq:pdf_rich}. 
The DM accurately reproduces the full distribution, including the rare, rapidly separating events. 
Crucially, deviations from the ideal Richardson self-similar stretched-exponential prediction are faithfully captured. 
The reconstruction of these deviations demonstrates that the model implicitly learns effective scale-dependent and history-dependent transport properties directly from data, without imposing any prior physical constraints. 
Moreover, when generating datasets larger than the DNS training set, the DM produces extended PDF tails corresponding to more extreme events than those observed during training, similarly to what was reported in~\cite{li2024synthetic}. 
This suggests that the model does not merely memorize the dataset but instead learns an approximation of the underlying data distribution, from which the generative process can sample new, physically consistent realizations.

\begin{figure}[t]
\includegraphics[width=1.\columnwidth]{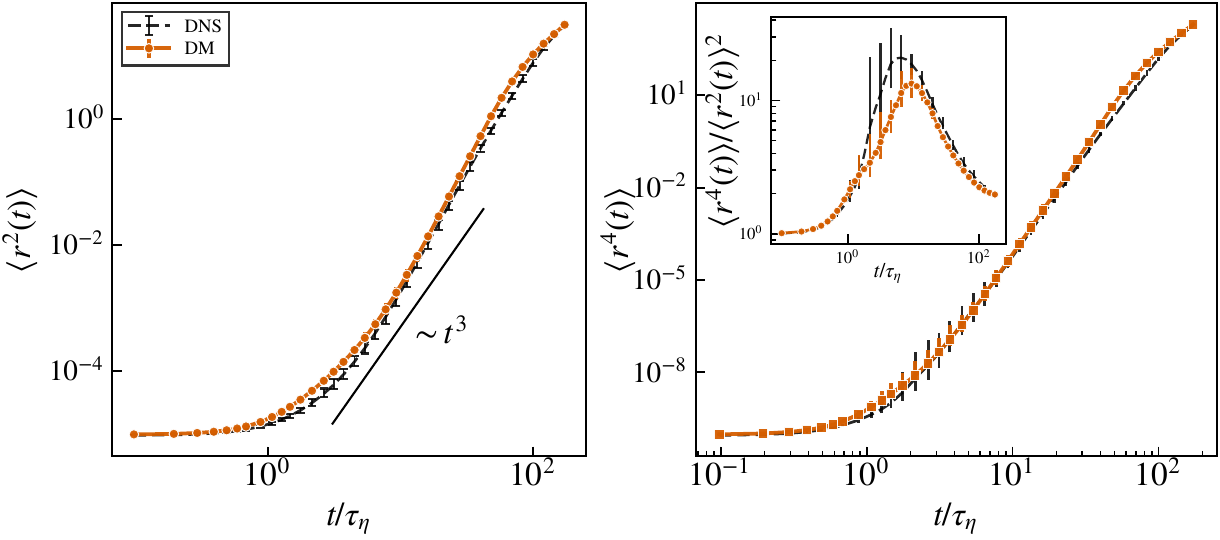}
\caption{Pair separation statistics as a function of time $t$ in units of $\tau_\eta$. 
Left: mean-square separation $\langle r^2(t) \rangle$, showing inertial-range scaling. The Richardson’s $t^3$ prediction is reported for reference. 
Right: fourth-order moment $\langle r^4(t) \rangle$. 
Right (inset): Flatness of the pair separation statistics, $\langle r^4(t) \rangle / \langle r^2(t) \rangle^2$. 
In all panels, DM (solid orange lines with full circles) is compared against DNS (dashed black lines).}
\label{fig:fig_3}
\end{figure}

To further quantify the agreement of the pair dispersion statistics, we analyze the moments of the inter-particle separation,
\begin{equation}
    \langle r^p(t) \rangle,
\end{equation}
where the average, $\langle \cdot \rangle$, is taken over all trajectory realizations and $p$ denotes the order of the moment. 
The instantaneous separation $r(t)$ is obtained by integrating the relative velocity between the two particles. 
Denoting by $\bm{V}^1(t)$ and $\bm{V}^2(t)$ the three-dimensional velocities of the two particles, the relative position evolves as
\begin{equation}
\bm{r}(t) = \bm{r}(t_0) + \int_0^t \big(\bm{V}^1(s)-\bm{V}^2(s)\big)\,ds,
\quad
r(t)=\|\bm{r}(t)\|.
\end{equation}
where $\bm{r}(t_0)$ is the initial separation vector used in the DNS, corresponding to half a grid spacing along one of the spatial directions ($x$, $y$, or $z$).
Fig.~\ref{fig:fig_3} reports the second- and fourth-order moments of pair separation averaged over all trajectories. 
The mean-square separation $\langle r^2(t) \rangle$ exhibits the expected super-diffusive growth compatible with Richardson’s $t^3$ scaling in the inertial range, in agreement with DNS. 
Higher-order statistics, including $\langle r^4(t) \rangle$, are also accurately reproduced. 
A particularly stringent measure for comparing intermittency in the two datasets is the flatness $\langle r^4(t) \rangle / \langle r^2(t) \rangle^2$. 
Unlike individual moments, which span several orders of magnitude, the flatness evolves within a narrow range and is therefore highlights statistical discrepancies. 
For self-similar (non-intermittent) statistics, this ratio would remain constant in time; its non-trivial behavior reflects the breakdown of simple scaling. 
The close agreement between DNS and DM demonstrates that the model captures intermittency corrections, including extreme and rare dispersion events, and not only the average or large scale behaviors.

\begin{figure}
\includegraphics[width=1.\columnwidth]{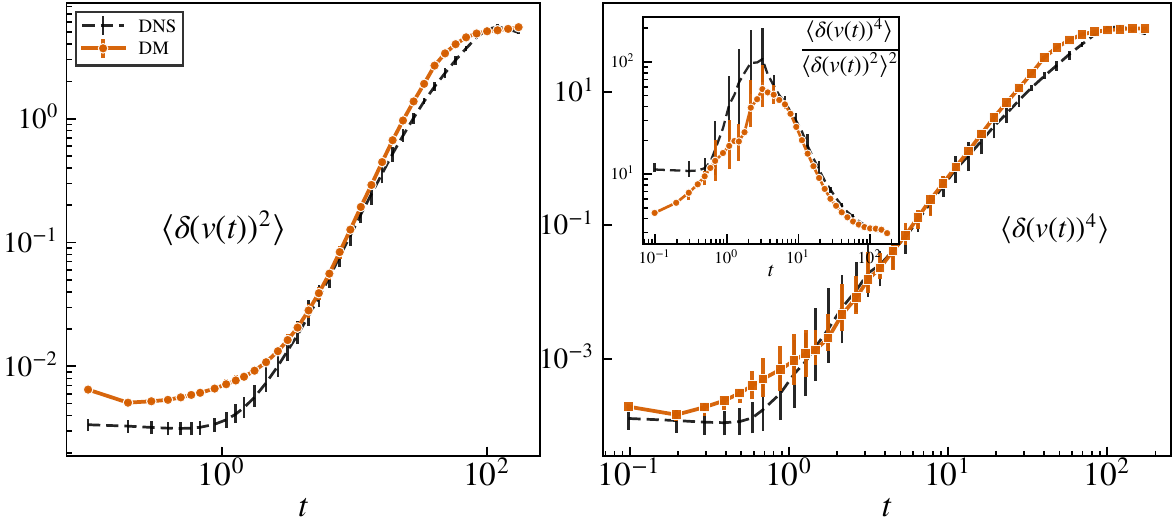}
\caption{Moments of the two particles velocity difference at the time lag $\tau$. 
Left: second-order moment $\langle (\delta v(\tau))^2 \rangle$. 
Right: fourth-order moment $\langle (\delta v(\tau))^4 \rangle$. 
Right (inset): Flatness $\langle (\delta v(\tau))^4 \rangle / \langle (\delta v(\tau))^2 \rangle^2$. In all panels, DM (solid orange lines with full circles) is compared against DNS (dashed black lines).}
\label{fig:fig_4}
\end{figure}

To investigate the statistics of the relative velocity of particles in the pair, we analyze their velocity difference at time $t$ during their evolution. 
For each pair realization, the relative velocity is defined as
\begin{equation}
    \delta v_i(t) = V_i^1(t) - V_i^2(t),
\end{equation}
where $V_i^1(t)$ and $V_i^2(t)$, with $i=x,y,z$, denote respectively the three velocity components of the first and the second particle in the pair at time $t$.
The $p$-th order moment of the relative velocity is then computed using isotropy by averaging the moments over the velocity components,
\begin{equation}
    \left\langle (\delta v(t))^{p} \right\rangle
    = \left\langle \frac{1}{3}
    \left( \delta v_x^{\,p} + \delta v_y^{\,p} + \delta v_z^{\,p} \right)
    \right\rangle,
\end{equation}
where $\langle \cdot \rangle$ is the average over an ensemble of different pairs. 
\noindent
As for the the pair dispersion, Figure~\ref{fig:fig_4} shows the second- and fourth-order moments of the relative velocities as functions of time, together with the associated flatness. 
The correct time dependence of such quantity indicates that the model encodes the non-trivial interplay between relative dispersion and multiscale velocity intermittency, a feature that cannot be captured by simple Markovian models based on local closures.

\begin{figure}
\includegraphics[width=1.\columnwidth]{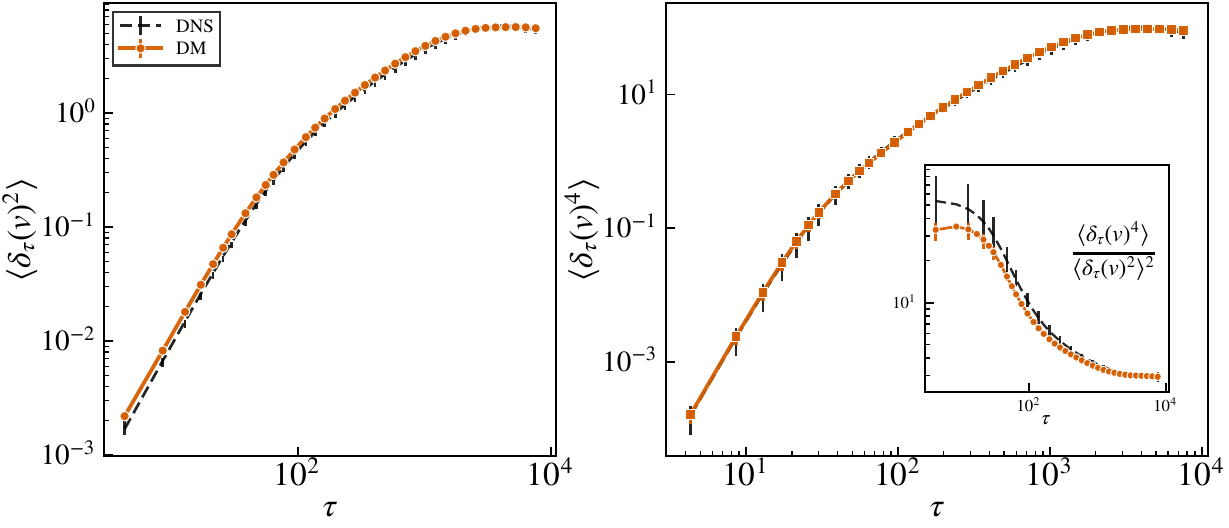}
\caption{Single-particle Lagrangian velocity increment statistics as a function of time lag $\tau$. 
Left: second-order structure function $\langle (\delta_\tau v)^2 \rangle$. 
Right: fourth-order structure function $\langle (\delta_\tau v)^4 \rangle$. 
Right (inset): flatness $\langle (\delta_\tau v)^4 \rangle / \langle (\delta_\tau v)^2 \rangle^2$. 
In all panels, DM (solid orange lines with full circles) is compared against DNS (dashed black lines).}
\label{fig:fig_5}
\end{figure}

\noindent
Importantly, the joint generation of particle pairs does not degrade single-particle statistics. 
To quantify single-particle Lagrangian fluctuations, we consider the velocity increments at time lag $\tau$,
\begin{equation}
    \delta_\tau V_i(t) = V_i(t+\tau) - V_i(t),
    \qquad i=x,y,z,
\end{equation}
and define the $p$-th order Lagrangian structure functions as
\begin{equation}
    S_\tau^{(p)} \coloneqq \left\langle \left(\delta_\tau V_i(t)\right)^p \right\rangle,
    \qquad p=2,4,
\end{equation}
where $\langle \cdot \rangle$ denotes an average over time $t$, over the three velocity components, and over all trajectory realizations. 
Figure~\ref{fig:fig_5} presents the second- and fourth-order structure functions as functions of the time lag $\tau$, showing excellent agreement with DNS across all resolved temporal scales. 
The growth of flatness at small $\tau$, reflecting intense acceleration events and strong temporal intermittency, is well preserved, with results comparable to those reported in~\cite{li2024synthetic} for single-particle statistics. 
These results demonstrate that the model simultaneously reproduces individual Lagrangian dynamics and realistic two-particle dispersion within a unified generative framework.

\section{Conclusion}
In this work, we addressed the long-standing problem of turbulent pair dispersion from a fully data-driven perspective. 
Our goal was to construct a stochastic generative framework capable of reproducing the joint dynamics of Lagrangian particle pairs while simultaneously preserving correct single-particle statistics. 
To this end, building on the recent success of denoising diffusion probabilistic models in generating turbulent data~\cite{li2024synthetic}, we extend this approach to the joint generation of two-particle velocity trajectories, training the model directly on high-resolution DNS data of homogeneous isotropic turbulence.
\noindent
The resulting DM accurately reproduces the multiscale evolution of pair separation, including deviations from Richardson’s classical $t^3$ scaling and from the self-similar stretched-exponential prediction for the separation PDF. 
Higher-order moments and flatness evolution confirm that intermittency corrections and rare extreme dispersion events are correctly captured. 
At the same time, the flatness of the single-particle Lagrangian structure functions remains in excellent agreement with DNS, demonstrating that the joint generation does not degrade individual particle dynamics.
\noindent
Together, these results demonstrate that diffusion models provide a fully data-driven, high-dimensional stochastic generative framework for turbulent single-particle and pair-dispersion dynamics. 
Without imposing Richardson scaling, eddy-diffusivity closures, or Markovian assumptions, the model learns effective multiscale and history-dependent transport properties directly from data. 
This establishes generative diffusion models as a new class of physics-consistent stochastic simulators capable of emulating non-Markovian, intermittent turbulent transport processes.

\section*{Acknowledgments}
This work was supported by the European Research Council (ERC) under the European Union’s Horizon 2020 research and innovation programme Smart-TURB (Grant Agreement No. 882340) and by the Italian Ministry of University and Research (MUR) - Fondo Italiano per la Scienza (FIS2) - 2023 Call, project DeepFL, CUP: E53C24003760001. This work was supported by the ANR grant SCALP  (ANR-24-CE23-1320).

\bibliographystyle{unsrt}
\bibliography{biblio}

@article{espanol2025effect,
  title={Effect of local flow geometry on particle pair dispersion angle},
  author={Espa{\~n}ol, Bernardo Luciano and Noseda, Manuel and Cobelli, Pablo Javier and Mininni, Pablo Daniel},
  journal={Physical Review Fluids},
  volume={10},
  number={4},
  pages={044501},
  year={2025},
  publisher={APS}
}

@article{shnapp2023universal,
  title={Universal alignment in turbulent pair dispersion},
  author={Shnapp, Ron and Brizzolara, Stefano and Neamtu-Halic, Marius M and Gambino, Alessandro and Holzner, Markus},
  journal={Nature Communications},
  volume={14},
  number={1},
  pages={4195},
  year={2023},
  publisher={Nature Publishing Group UK London}
}

@article{sreenivasan2019turbulent,
  title={Turbulent mixing: A perspective},
  author={Sreenivasan, Katepalli R},
  journal={Proceedings of the National Academy of Sciences},
  volume={116},
  number={37},
  pages={18175--18183},
  year={2019},
  publisher={National Academy of Sciences}
}

@article{boffetta2002statistics,
  title={Statistics of two-particle dispersion in two-dimensional turbulence},
  author={Boffetta, Guido and Sokolov, Igor M},
  journal={Physics of fluids},
  volume={14},
  number={9},
  pages={3224--3232},
  year={2002},
  publisher={American Institute of Physics}
}

@article{bourgoin2015turbulent,
  title={Turbulent pair dispersion as a ballistic cascade phenomenology},
  author={Bourgoin, Micka{\"e}l},
  journal={Journal of Fluid Mechanics},
  volume={772},
  pages={678--704},
  year={2015},
  publisher={Cambridge University Press}
}

@article{mathai2018dispersion,
  title={Dispersion of air bubbles in isotropic turbulence},
  author={Mathai, Varghese and Huisman, Sander G and Sun, Chao and Lohse, Detlef and Bourgoin, Micka{\"e}l},
  journal={Physical review letters},
  volume={121},
  number={5},
  pages={054501},
  year={2018},
  publisher={APS}
}

@article{grossmann1984unified,
  title={Unified theory of relative turbulent diffusion},
  author={Grossmann, Siegfried and Procaccia, Itamar},
  journal={Physical Review A},
  volume={29},
  number={3},
  pages={1358},
  year={1984},
  publisher={APS}
}

@article{crisanti1987multifractal,
  title={Is multifractal structure relevant for the turbulent diffusion?},
  author={Crisanti, Andrea and Paladin, G and Vulpiani, Angelo},
  journal={Physics Letters A},
  volume={126},
  number={2},
  pages={120--122},
  year={1987},
  publisher={Elsevier}
}

@article{novikov1989two,
  title={Two-particle description of turbulence, Markov property, and intermittency},
  author={Novikov, EA},
  journal={Physics of Fluids A: Fluid Dynamics},
  volume={1},
  number={2},
  pages={326--330},
  year={1989},
  publisher={American Institute of Physics}
}

@article{majda1999simplified,
  title={Simplified models for turbulent diffusion: theory, numerical modelling, and physical phenomena},
  author={Majda, Andrew J and Kramer, Peter R},
  journal={Physics reports},
  volume={314},
  number={4-5},
  pages={237--574},
  year={1999},
  publisher={Elsevier}
}

@article{zouari1994derivation,
  title={Derivation of the relative dispersion law in the inverse energy cascade of two-dimensional turbulence},
  author={Zouari, Najet and Babiano, Armando},
  journal={Physica D: Nonlinear Phenomena},
  volume={76},
  number={1-3},
  pages={318--328},
  year={1994},
  publisher={Elsevier}
}

@article{pope2001turbulent,
  title={Turbulent flows},
  author={Pope, Stephen B},
  journal={Measurement Science and Technology},
  volume={12},
  number={11},
  pages={2020--2021},
  year={2001}
}

@book{davidson2011voyage,
  title={A voyage through turbulence},
  author={Davidson, Peter A and Kaneda, Yukio and Moffatt, Keith and Sreenivasan, Katepalli R},
  year={2011},
  publisher={Cambridge University Press}
}

@article{scatamacchia2012extreme,
  title={Extreme Events in the Dispersions of Two Neighboring Particles Under<? format?> the Influence of Fluid Turbulence},
  author={Scatamacchia, R and Biferale, L and Toschi, F},
  journal={Physical review letters},
  volume={109},
  number={14},
  pages={144501},
  year={2012},
  publisher={APS}
}

@article{bourgoin2006role,
  title={The role of pair dispersion in turbulent flow},
  author={Bourgoin, Micka{\"e}l and Ouellette, Nicholas T and Xu, Haitao and Berg, Jacob and Bodenschatz, Eberhard},
  journal={Science},
  volume={311},
  number={5762},
  pages={835--838},
  year={2006},
  publisher={American Association for the Advancement of Science}
}

@article{salazar2009two,
  title={Two-particle dispersion in isotropic turbulent flows},
  author={Salazar, Juan PLC and Collins, Lance R},
  journal={Annual review of fluid mechanics},
  volume={41},
  number={1},
  pages={405--432},
  year={2009},
  publisher={Annual Reviews}
}

@book{frisch1996turbulence,
  title={Turbulence: the legacy of AN Kolmogorov},
  author={Frisch, Uriel},
  year={1996},
  publisher={CUP}
}

@book{monin2013statistical,
  title={Statistical fluid mechanics, volume II: mechanics of turbulence},
  author={Monin, Andre{i} Sergeevich and Yaglom, Akiva M},
  volume={2},
  year={2013},
  publisher={Dover}
}

@article{boffetta2002relative,
  title={Relative dispersion in fully developed turbulence: the Richardson’s law and intermittency corrections},
  author={Boffetta, Guido and Sokolov, Igor M},
  journal={Physical review letters},
  volume={88},
  number={9},
  pages={094501},
  year={2002},
  publisher={APS}
}

@article{boffetta1999pair,
  title={Pair dispersion in synthetic fully developed turbulence},
  author={Boffetta, Guido and Celani, A and Crisanti, Andrea and Vulpiani, Angelo},
  journal={Physical Review E},
  volume={60},
  number={6},
  pages={6734},
  year={1999},
  publisher={APS}
}

@article{falkovich2001particles,
  title={Particles and fields in fluid turbulence},
  author={Falkovich, Gregory and Gawedzki, K and Vergassola, Massimo},
  journal={Reviews of modern Physics},
  volume={73},
  number={4},
  pages={913},
  year={2001},
  publisher={APS}
}

@article{sawford2001turbulent,
  title={Turbulent relative dispersion},
  author={Sawford, Brian},
  journal={Annual review of fluid mechanics},
  volume={33},
  number={1},
  pages={289--317},
  year={2001},
  publisher={Annual Reviews 4139 El Camino Way, PO Box 10139, Palo Alto, CA 94303-0139, USA}
}

@article{richardson1926atmospheric,
  title={Atmospheric diffusion shown on a distance-neighbour graph},
  author={Richardson, Lewis Fry},
  journal={Proceedings of the Royal Society of London. Series A, Containing Papers of a Mathematical and Physical Character},
  volume={110},
  number={756},
  pages={709--737},
  year={1926},
  publisher={The Royal Society London}
}

@article{li2024synthetic,
  title={Synthetic Lagrangian turbulence by generative diffusion models},
  author={Li, Tianyi and Biferale, Luca and Bonaccorso, Fabio and Scarpolini, Martino Andrea and Buzzicotti, Michele},
  journal={Nature Machine Intelligence},
  volume={6},
  number={4},
  pages={393--403},
  year={2024},
  publisher={Nature Publishing Group UK London}
}

@inproceedings{sohl2015deep,
  title={Deep unsupervised learning using nonequilibrium thermodynamics},
  author={Sohl-Dickstein, Jascha and Weiss, Eric and Maheswaranathan, Niru and Ganguli, Surya},
  booktitle={International conference on machine learning},
  pages={2256--2265},
  year={2015},
  organization={pmlr}
}

@article{ho2020denoising,
  title={Denoising diffusion probabilistic models},
  author={Ho, Jonathan and Jain, Ajay and Abbeel, Pieter},
  journal={Advances in neural information processing systems},
  volume={33},
  pages={6840--6851},
  year={2020}
}

@article{li2024generative,
  title={Generative diffusion models for synthetic trajectories of heavy and light particles in turbulence},
  author={Li, Tianyi and Tommasi, Samuele and Buzzicotti, Michele and Bonaccorso, Fabio and Biferale, Luca},
  journal={International Journal of Multiphase Flow},
  volume={181},
  pages={104980},
  year={2024},
  publisher={Elsevier}
}

@article{martin2025generation,
  title={Generation of Cosmic-Ray Trajectories by a Diffusion Model Trained on Test Particles in 3D Magnetohydrodynamic Turbulence},
  author={Martin, Johannes and L{\"u}bke, Jeremiah and Li, Tianyi and Buzzicotti, Michele and Grauer, Rainer and Biferale, Luca},
  journal={The Astrophysical Journal Supplement Series},
  volume={277},
  number={2},
  pages={48},
  year={2025},
  publisher={IOP Publishing}
}

@article{li2025stochastic,
  title={Stochastic reconstruction of gappy Lagrangian turbulent signals by conditional diffusion models},
  author={Li, Tianyi and Biferale, Luca and Bonaccorso, Fabio and Buzzicotti, Michele and Centurioni, Luca},
  journal={Communications Physics},
  volume={8},
  number={1},
  pages={372},
  year={2025},
  publisher={Nature Publishing Group UK London}
}

@article{biferale2023turb,
  title={Turb-lagr. a database of 3d lagrangian trajectories in homogeneous and isotropic turbulence},
  author={Biferale, Luca and Bonaccorso, Fabio and Buzzicotti, Michele and Calascibetta, Chiara},
  journal={arXiv preprint arXiv:2303.08662},
  year={2023}
}

@article{calascibetta2023taming,
  title={Taming Lagrangian chaos with multi-objective reinforcement learning},
  author={Calascibetta, Chiara and Biferale, Luca and Borra, Francesco and Celani, Antonio and Cencini, Massimo},
  journal={The European Physical Journal E},
  volume={46},
  number={3},
  pages={9},
  year={2023},
  publisher={Springer}
}

@article{calascibetta2023optimal,
  title={Optimal tracking strategies in a turbulent flow},
  author={Calascibetta, Chiara and Biferale, Luca and Borra, Francesco and Celani, Antonio and Cencini, Massimo},
  journal={Communications Physics},
  volume={6},
  number={1},
  pages={256},
  year={2023},
  publisher={Nature Publishing Group UK London}
}

@article{toschi2009lagrangian,
  title={Lagrangian properties of particles in turbulence},
  author={Toschi, Federico and Bodenschatz, Eberhard},
  journal={Annual review of fluid mechanics},
  volume={41},
  number={1},
  pages={375--404},
  year={2009},
  publisher={Annual Reviews}
}

@article{li2024relative,
  title={Relative dispersion in free-surface turbulence},
  author={Li, Yaxing and Wang, Yifan and Qi, Yinghe and Coletti, Filippo},
  journal={Journal of Fluid Mechanics},
  volume={993},
  pages={R2},
  year={2024},
  publisher={Cambridge University Press}
}

@article{buzzicotti2023data,
  title={Data reconstruction for complex flows using AI: Recent progress, obstacles, and perspectives},
  author={Buzzicotti, Michele},
  journal={Europhysics Letters},
  volume={142},
  number={2},
  pages={23001},
  year={2023},
  publisher={EDP Sciences, IOP Publishing and Societ{\`a} Italiana di Fisica}
}

@article{bec2010turbulent,
  title={Turbulent pair dispersion of inertial particles},
  author={Bec, J and Biferale, L and Lanotte, AS and Scagliarini, Andrea and Toschi, F},
  journal={Journal of Fluid Mechanics},
  volume={645},
  pages={497--528},
  year={2010},
  publisher={Cambridge University Press}
}

\end{document}